\documentclass[12pt]{article}
\usepackage{amsfonts}
\usepackage{color}
\usepackage{footmisc}
\usepackage[numbers,sort&compress]{natbib}
\usepackage[bookmarksopen=true, bookmarksopenlevel=\maxdimen,
bookmarksnumbered=true,colorlinks,linkcolor=blue,
citecolor=blue,urlcolor=blue]{hyperref}

\newcommand{\omits}[1]{}

\definecolor{dyellow}{rgb}{1.,0.8,.0}
\definecolor{myblue}{rgb}{.1,.1,.7}
\definecolor{dcyan}{rgb}{.0,.6,.6}
\definecolor{dmagenta}{rgb}{0.6,0.0,0.6}
\definecolor{brown}{rgb}{0.6,0.2,0.}
\definecolor{darkblue}{rgb}{.0,.0,0.5}
\definecolor{darkred}{rgb}{0.75,0.0,0.0}
\definecolor{orange}{rgb}{1.,.6,.0}
\definecolor{dorange}{rgb}{0.8,.4,.0}
\definecolor{darkgreen}{rgb}{0.0,0.6,0.0}
\definecolor{purple}{rgb}{.4,.0,.4}
\definecolor{lightgrey}{rgb}{0.7,0.7,0.7}

\begin{document}
\phantomsection \addcontentsline{toc}{chapter}{Weak field
approximation in a model of de Sitter gravity: Schwarzschild
solutions and galactic rotation curves}
\begin{center}
{\bf \LARGE Weak field approximation in\\ a model of de Sitter gravity:\\
Schwarzschild solutions and\bigskip\\ galactic rotation curves}

\bigskip

\bigskip

{\Large Jia-An Lu$^{a}$\footnote{Email: ljagdgz@163.com} and
Chao-Guang Huang$^{b}$\footnote{Email: huangcg@ihep.ac.cn} }

\bigskip

$^a$\ School of Mathematics and Computational Science, Sun Yat-sen
University, Guangzhou 510275, China

$^b$\ Institute of High Energy Physics, and
Theoretical Physics Center for \\
Science Facilities, Chinese Academy of Sciences, Beijing 100049,
China

\setcounter{footnote}{-1} \footnote{The final publication is
available at
\href{http://www.springerlink.com/openurl.asp?genre=article&id=doi:10.1007/s10714-012-1494-5}
{www.springerlink.com}.}

\phantomsection \addcontentsline{toc}{section}{Abstract}
\begin{abstract}
Weak field approximate solutions in the $\Lambda\rightarrow 0$ limit
of a model of de Sitter gravity have been presented in the static
and spherically symmetric case. Although the model looks different
from general relativity, among those solutions, there still exist
the weak Schwarzschild fields with the smooth connection to regular
internal solutions obeying the Newtonian gravitational law. The
existence of such solutions would determine the value of the
coupling constant, which is different from that of the previous
literature. Moreover, there also exist solutions that could deduce
the galactic rotation curves without invoking dark matter.
\end{abstract}
\end{center}


\section{Introduction}
In the 1970s a model of de Sitter (dS) gravity had been proposed
\cite{{Wu},{An},{Guo79},{Townsend}}. In this model the
Einstein--Hilbert action with a cosmological term could be deduced
from a gauge-like action besides two quadratic terms of the
curvature and torsion. The astronomical observation
\cite{{Bennett},{Tegmark}} on the asymptotically dS behavior of our
universe has increased interest in the model as it may offer a way
to deal with the dark energy problem \cite{Peebles}. If the
Einstein--Hilbert term is required to be the main part of the
gauge-like action, the cosmological constant should be large. The
large cosmological constant may be canceled out by the vacuum energy
density, leaving a small cosmological constant \cite{Townsend}. But
it is difficult to explain why the large cosmological constant and
the vacuum energy density are so close, but not exactly equal, to
each other. On the other hand, if the cosmological constant is
required to be small \cite{{Wu},{An},{Guo79},{Guo07}}, the quadratic
curvature term would become the main part of the action. Note that
in one of the field equations, the quadratic curvature term only
contributes to the symmetric trace-free part. It is worth checking
carefully whether the model under this case could explain the
experimental observations. Actually it has been shown
\cite{{cos1},{cos2}} that the model with a small cosmological
constant may explain the accelerating expansion of the universe and
supply a natural transit from decelerating expansion to accelerating
expansion without the help of introducing matter fields in addition
to dust. It has also been shown \cite{{vacuum1},{vacuum2}} that all
torsion-free vacuum solutions of this model are the vacuum solutions
of Einstein's field equation with the same cosmological constant,
and vise versa. Therefore, one may expect that the model with a
small cosmological constant may pass all solar-system-scale
experimental tests for general relativity (GR).

However, it has been pointed out \cite{cos1} that the
energy-momentum-stress tensor of a spinless fluid in the
torsion-free case of this model should be with a constant trace.
Questions would then appear such as, could the torsion-free
Schwarzschild-dS (S-dS) solution be smoothly connected to internal
solutions with nonzero torsion, or are there any S-dS solutions with
nonzero torsion? In fact, the different dS spacetimes with nonzero
torsion in this model have been obtained in
\cite{{dStor1},{dStor2}}, but they are sill not the S-dS solutions.
On the other hand, S-dS solutions with long-range spherically
symmetric torsion have been given \cite{Chen} in some special cases
(not necessarily under the double duality ansatz \cite{Lenzen}) of
quadratic models of Poincar\'e gauge theory of gravity, but our
model does not fall into those special cases.

We would like to firstly check the existence of the S-dS solutions
with nonzero torsion in the weak field approximation. The Newtonian
limit of general quadratic models in Poincar\'e gauge theory of
gravity has been calculated \cite{{Hayashi3},{Zhang83}} in the
1980's. In those calculations, quadratic terms in the field
equations have been thrown away as usual. However, in the weak field
approximation of our model, the quadratic curvature terms could not
be easily thrown away, for the reason that they are the main parts
of one of the field equations. In fact, we may let the cosmological
constant be $\Lambda\rightarrow 0$, then only those quadratic
curvature terms would appear in the limit of the field equation
which contains the energy-momentum-stress tensor of the matter
field. On the other hand, as those quadratic curvature terms are
symmetric and trace-free, they would not appear in the trace part
and the antisymmetric part of the field equations. The Newtonian
limit of the trace equations has been recently analyzed \cite{Ma},
but a more complete analysis of all components of the field
equations is needed.

As was well known, Newton's theory of gravity meets great
difficulties in the explanation of the flat rotation curves
\cite{RC} of spiral galaxies. The most widely adopted way to resolve
this problem is the dark matter hypothesis. But up to now, all of
the possible candidates of dark matter (such as neutralino, axion,
etc.) are either undetected or unsatisfactory. In the meanwhile,
there also exist some models
\cite{MOND1,MOND2,higher-order-grav,finsler} which could deduce the
galactic rotation curves without involving dark matter. We would
like to explore the possibility of a new explanation for the
galactic rotation curves from the dS gravity model.

The paper is arranged as follows. We first briefly review the model
of the dS gravity in section 2. In the third section, after dividing
a field equation into its trace part, symmetric trace-free part and
antisymmetric part, we attain the $\Lambda\rightarrow 0$ limit of
the model and calculate its weak field approximation in the static
and spherically symmetric case. The weak field approximate solutions
contain the weak Schwarzschild fields with nonzero torsion, which
could be smoothly linked to regular internal solutions obeying the
Newtonian gravitational law. The coupling constant is determined by
the existence of such solutions. Moreover, solutions that could
deduce the galactic rotation curves without invoking dark matter are
also attained. Finally we end with some remarks in the last section.

\section{A model of dS gravity}
A model of dS gravity has been constructed with a gauge-like action
\cite{Wu,An,Guo79,Townsend}
\begin{eqnarray}\label{Action}
S_{G}&=&\int L_{G}=\int \kappa[-\textrm{tr}(\mathcal
{F}_{ab}\mathcal {F}^{ab})]
\nonumber\\
&=&\int\kappa[R_{abcd}R^{abcd}-\frac{4}{l^{2}}(R-\frac{6}{l^{2}})
+\frac{2}{l^{2}}S_{abc}S^{abc}]
\end{eqnarray}
in the units of $\hbar=c=1$, where $\kappa$ is a dimensionless
coupling constant to be determined, and
\begin{equation}
\mathcal {F}_{ab}=(d\mathcal {A}+\frac{1}{2}[\mathcal {A},\mathcal
{A}])_{ab}
\end{equation}
or explicitly
\begin{eqnarray}
\mathcal {F}^{A}{}_{Bab}&=&(d \mathcal {A}^{A}{}_{B})_{ab}+\mathcal
{A}^{A}{}_{Ca}\wedge\mathcal {A}^{C}{}_{Bb}
\nonumber\\
&=&\left(
\begin{array}{cc}
R_{ab}{}^{\alpha}{}_{\beta}-l^{-2}e^{\alpha}{}_{a}\wedge e_{\beta b}
&l^{-1}S^{\alpha}{}_{ab}\\
-l^{-1}S_{\beta ab}&0
\end{array}
\right)
\end{eqnarray}
is a dS algebra-valued 2-form and
\begin{equation}
\mathcal {A}^{A}{}_{Ba}=\left(
\begin{array}{cc}
\Gamma^{\alpha}{}_{\beta a}&l^{-1}e^{\alpha}{}_{a}\\
-l^{-1}e_{\beta a}&0
\end{array}
\right)
\end{equation}
is a dS algebra-valued 1-form. Here $A,B...=0,1,2,3,4$ stand for
matrix indices (internal indices) and the trace in Eq.
(\ref{Action}) is taken for those indices. In addition,
$\{e_{\alpha}{}^{a}\}$ is some local orthonormal frame field on the
spacetime manifold and $\Gamma^{\alpha}{}_{\beta a}$ is the
connection 1-form in this frame field, where $a,b...$ stand for
abstract indices \cite{{Liang},{Wald}} and $\alpha,\beta...=0,1,2,3$
are concrete indices related to the frame field mentioned above. The
curvature 2-form $R_{ab}{}^{\alpha}{}_{\beta}$ and torsion 2-form
$S^{\alpha}{}_{ab}$ are related to the connection 1-form
$\Gamma^{\alpha}{}_{\beta a}$ as follows:
\begin{equation}
R_{ab}{}^{\alpha}{}_{\beta}=(d\Gamma^{\alpha}{}_{\beta})_{ab}+\Gamma^{\alpha}{}_{\gamma
a}\wedge\Gamma^{\gamma}{}_{\beta b},
\end{equation}
\begin{equation}
S^{\alpha}{}_{ab}=(de^{\alpha})_{ab}+\Gamma^{\alpha}{}_{\beta
a}\wedge e^{\beta}{}_{b}.
\end{equation}
Moreover,
\begin{displaymath}
R_{abc}{}^{d}=R_{ab\alpha}{}^{\beta}e^{\alpha}{}_{c}e_{\beta}{}^{d},\quad
S^{c}{}_{ab}=S^{\alpha}{}_{ab}e_{\alpha}{}^{c},
\end{displaymath}
\begin{displaymath}
R_{ab}=R_{acb}{}^{c}, \quad R=g^{ab}R_{ab}.
\end{displaymath}
In fact, if spacetime is an umbilical submanifold of some
(1+4)-dimensional ambient manifold and with positive normal
curvature, then $\mathcal {A}_{a}$ and $\mathcal {F}_{ab}$ could be
viewed \cite{Guo07} as the connection 1-form and curvature 2-form
(in the dS-Lorentz frame) of the ambient manifold restricted to
spacetime. Here, an umbilical submanifold means a submanifold with
constant normal curvature, such as the dS spacetime which could be
seen as an umbilical submanifold of a 5d Minkowski spacetime with
positive normal curvature. $\mathcal {A}_{a}$ could also be seen
\cite{Wise} as the Cartan connection of a Cartan geometry modeled on
the dS spacetime and based on the spacetime manifold, with $\mathcal
{F}_{ab}$ the corresponding curvature 2-form. The Cartan geometry is
a generalization of homogenous spaces with fibre bundle language,
and one may refer to \cite{Wise} for more details. In addition, it
should be noted that $3/l^{2}$ is identified \cite{Guo07} with a
small cosmological constant $\Lambda$ here, which is very different
from the viewpoint of \cite{Townsend} where $l$ is identified with
the Planck length. The signature is chosen such that the metric
coefficients are $\eta_{\alpha\beta}=\textrm{diag}(-1,1,1,1)$.

The total action is $S=S_{M}+S_{G}$, where $S_{M}$ is the action of
the matter fields and the field equations can be given via the
variational principle with respect to
$e^{\alpha}{}_{a},~\Gamma^{\alpha}{}_{\beta a}$:
\begin{eqnarray}\label{1stEq}
\frac{8}{l^{2}}(G_{ab}+\Lambda g_{ab})+|R|^{2}g_{ab}
-4R_{acde}R_{b}{}^{cde}+\frac{2}{l^{2}}|S|^{2}g_{ab}
\nonumber\\
-\frac{8}{l^{2}}S_{cda}S^{cd}{}_{b}
+\frac{8}{l^{2}}\nabla_{c}S_{ab}{}^{c}
+\frac{4}{l^{2}}S_{acd}T_{b}{}^{cd}+\frac{1}{\kappa}\Sigma_{ab}=0,
\end{eqnarray}
\begin{equation}\label{2ndEq}
-\frac{4}{l^{2}}T^{a}{}_{bc}-4\nabla_{d}R^{da}{}_{bc}
+2T^{a}{}_{de}R^{de}{}_{bc}
-\frac{8}{l^{2}}S_{[bc]}{}^{a}+\frac{1}{\kappa}\tau_{bc}{}^{a}=0,
\end{equation}
where
\begin{displaymath}
G_{ab}=R_{ab}-\frac{1}{2}Rg_{ab}, \quad
T^{c}{}_{ab}=S^{c}{}_{ab}+2\delta^{c}{}_{[a}S^{d}{}_{b]d},
\end{displaymath}
\begin{displaymath}
|R|^{2}=R_{abcd}R^{abcd}, \quad |S|^{2}=S_{abc}S^{abc},
\end{displaymath}
\begin{displaymath}
\Sigma_{\alpha}{}^{a}=\delta S_{M}/\delta e^{\alpha}{}_{a},\quad
\Sigma_{b}{}^{a}=\Sigma_{\alpha}{}^{a}e^{\alpha}{}_{b},
\end{displaymath}
\begin{displaymath}
\tau_{\alpha}{}^{\beta a}=\delta
S_{M}/\delta\Gamma^{\alpha}{}_{\beta a},\quad
\tau_{b}{}^{ca}=\tau_{\alpha}{}^{\beta
a}e^{\alpha}{}_{b}e_{\beta}{}^{c},
\end{displaymath}
and the variational derivatives are defined as follows: if
\begin{displaymath}
\delta S_{M}=\int(X_{\alpha}{}^{a}\delta
e^{\alpha}{}_{a}+Y_{\alpha\beta}{}^{a}\delta
\Gamma^{\alpha\beta}{}_{a}),
\end{displaymath}
then
\begin{displaymath}
\delta S_{M}/\delta e^{\alpha}{}_{a}=X_{\alpha}{}^{a},\quad \delta
S_{M}/\delta\Gamma^{\alpha\beta}{}_{a}=Y_{[\alpha\beta]}{}^{a}.
\end{displaymath}

\section{Weak field approximation in the case with \texorpdfstring{$\Lambda\rightarrow 0$}{Lambda~-->~0}}
As $\Lambda$ is very small, it is interesting to see the case with
$\Lambda\rightarrow 0$ ($l\to\infty$). If $\Lambda\rightarrow 0$ is
directly set in the first field equation, then only the quadratic
curvature terms are left, which are symmetric and trace-free. Thus,
we would like to perform the following procedure. Divide the first
field equation into its symmetric trace-free part, trace part, and
antisymmetric part, then let $l$ tend to infinity
($l\rightarrow\infty$) in the above three parts and in the second
field equation. The limiting equations are:
\begin{equation}\label{sym-tracefreeEq}
|R|^{2}g_{ab} -4R_{acde}R_{b}{}^{cde}=0,
\end{equation}
\begin{equation}\label{traceEq}
-R-\nabla_{c}S^{bc}{}_{b}+\frac{1}{2}S_{bcd}T^{bcd}+(l^{2}/8\kappa)\Sigma=0,
\end{equation}
\begin{equation}\label{antisymEq}
R_{[ab]}+\nabla_{c}S_{[ab]}{}^{c}+\frac{1}{2}S_{[a}{}^{cd}T_{b]}{}_{cd}+(l^{2}/8\kappa)\Sigma_{[ab]}=0,
\end{equation}
\begin{equation}\label{2ndEqlimit}
-2\nabla_{d}R^{da}{}_{bc}+T^{a}{}_{de}R^{de}{}_{bc}=0.
\end{equation}
When $l\rightarrow\infty$, $l^{2}/\kappa$ should tend to a finite
value, otherwise Eqs. (\ref{traceEq}) and (\ref{antisymEq}) would
give $\Sigma=0$ and $\Sigma_{[ab]}=0$, which are unreasonable. In
the torsion-free case, the scalar curvature would be a constant from
Eq. (\ref{2ndEqlimit}), and, therefore, $\Sigma=$~const from Eq.
(\ref{traceEq}). This property has been pointed out by \cite{cos1}.
Now we are going to consider the week field approximation of the
above equations. It would be assumed that
\begin{equation}\label{weakmetric-torsion}
g_{ab}=\eta_{ab}+\gamma_{ab},\ \gamma_{ab}=O(s),\ S^{c}{}_{ab}=O(s),
\end{equation}
\begin{equation}\label{weakmatter}
\Sigma_{ab}=O(s),\quad \tau_{ab}{}^{c}=O(s),
\end{equation}
where $s$ is a dimensionless parameter, called the weak field
parameter. We will restrict ourselves to the static and
$O(3)$-symmetric case, with the static spherical coordinate system
$\{t,r,\theta,\varphi\}$. $\eta_{ab}$ could be defined by its
components $\eta_{\mu\nu}=\textrm{diag}(-1,1,1,1)$ in the
approximate inertial coordinate system $\{x^{\mu}\}$. $\{x^{\mu}\}$
is related to $\{t,r,\theta,\varphi\}$ as usual:
\begin{displaymath}
x^{0}=t,\ x^{1}=r\sin\theta\cos\varphi,\
x^{2}=r\sin\theta\sin\varphi,\ x^{3}=r\cos\theta.
\end{displaymath}
It could be proved \cite{Chen,Liang,Wheeler} that $\gamma_{ab}$ and
$S^{c}{}_{ab}$ have only these dependent components in the static
spherical coordinate system:
\begin{equation}\label{1stordermetric}
\gamma_{00}=-2\phi(r), \quad \gamma_{rr}=-2\psi(r),
\end{equation}
\begin{equation}\label{staticO3torsion}
\left\{\begin{array}{ll}S^{0}{}_{0r}=f(r), \quad S^{r}{}_{0r}=h(r),\\
S^{\theta}{}_{r\theta}=g(r), \quad S^{\theta}{}_{0\theta}=-k(r),\\
S^{\varphi}{}_{r\varphi}=g(r), \quad S^{\varphi}{}_{0\varphi}=-k(r),
\end{array}\right.
\end{equation}
where $\phi$ plays the role of the Newtonian gravitational potential
\cite{Wu,An,Guo79,Liang,Zhang85} and $\psi$ is an unknown function.
It can be shown that components of $\gamma_{ab}$ and $S^{c}{}_{ab}$
in $\{x^{\mu}\}$ are as follows:
\begin{equation}\label{1stordermetric2}
\gamma_{00}=-2\phi, \quad \gamma_{0i}=0, \quad
\gamma_{ij}=(-2\psi)x_{i}x_{j}/r^{2},
\end{equation}
\begin{equation}\label{staticO3torsion2}
\left\{\begin{array}{ll}S^{0}{}_{0i}=fx_{i}/r,\quad S^{0}{}_{ij}=0,\\
S^{i}{}_{0j}=(h+k)x^{i}x_{j}/r^{2}-k\delta^{i}{}_{j},\\
S^{i}{}_{jk}=(-g/r)(\delta^{i}{}_{j}x_{k}-\delta^{i}{}_{k}x_{j}).
\end{array}\right.
\end{equation}
For this case the contorsion tensor is related to the torsion tensor
by
\begin{equation}\label{Contorsion-Torsion}
K_{abc}=S_{cba}.
\end{equation}
Let
$\Gamma^{c}{}_{ab}=\Gamma^{\sigma}{}_{\mu\nu}\partial_{\sigma}{}^{c}(dx^{\mu})_{a}(dx^{\nu})_{b}$,
where $\Gamma^{\sigma}{}_{\mu\nu}$ is the connection coefficient in
$\{x^{\mu}\}$. $\Gamma^{c}{}_{ab}$ and the curvature tensor have the
following first order approximate expressions:
\begin{equation}\label{connection}
\Gamma^{c}{}_{ab}=\frac{1}{2}(\partial_{a}\gamma_{b}{}^{c}
+\partial_{b}\gamma_{a}{}^{c}-\partial^{c}\gamma_{ab})-K^{c}{}_{ab},
\end{equation}
\begin{equation}\label{Curvature}
R_{abc}{}^{d}=-(\partial_{c}\partial_{[a}\gamma_{b]}{}^{d}
-\partial^{d}\partial_{[a}\gamma_{b]c})+2\partial_{[a}K^{d}{}_{|c|b]},
\end{equation}
where
\begin{equation}\label{Contorsion}
K^{c}{}_{ab}=\frac{1}{2}(S^{c}{}_{ab}+S_{ab}{}^{c}+S_{ba}{}^{c})
\end{equation}
is the contorsion tensor. In the case with
$\Sigma_{ab}=\rho(r)(dt)_{a}(dt)_{b}$, the equations for the first
order approximate $g_{ab}$ and $S^c{}_{ab}$ are as follows:
\begin{equation}\label{Lambda0SymEq}
|R|^{2}\eta_{ab}-4R_{acde}R_{b}{}^{cde}=0,
\end{equation}
\begin{equation}\label{Lambda0TraceEq}
R+\partial_{c}S^{bc}{}_{b}+(l^{2}/8\kappa)\rho=0,
\end{equation}
\begin{equation}\label{Lambda0antisymEq}
R_{[ab]}+\partial_{c}S_{[ab]}{}^{c}=0,
\end{equation}
\begin{equation}\label{Lambda0-2ndEq}
\partial_{d}R^{da}{}_{bc}=0.
\end{equation}
The first order approximation of Eq. (\ref{sym-tracefreeEq}) is an
identity. Here Eq. (\ref{Lambda0SymEq}) is the second order
approximation of Eq. (\ref{sym-tracefreeEq}). It should be
considered for the reason that it is still an equation for the first
order approximate $g_{ab}$ and $S^c{}_{ab}$.

Now we are going to solve the above weak field equations. Applying
Eq. (\ref{Curvature}) to Eq. (\ref{Lambda0-2ndEq}), we have
\begin{equation}\label{Lambda0-2ndEq2}
\partial_{d}(\partial^{[d}\partial_{[c}\gamma_{b]}{}^{a]}+\partial^{[d}K_{cb}{}^{a]})=0.
\end{equation}
The $00i$ $(abc)$ component of Eq. (\ref{Lambda0-2ndEq2}) is
\begin{displaymath}
\partial_{d}(\partial^{d}\partial_{i}\gamma_{00}+2\partial^{d}K_{i00})=0,
\end{displaymath}
which gives
\begin{displaymath}
\triangle[(\phi'+f)x_{i}/r]=0,
\end{displaymath}
\begin{displaymath}
i.e.\quad (\phi'+f)''/r+2(\phi'+f)'/r^{2}-2(\phi'+f)/r^{3}=0.
\end{displaymath}
Solving this equation we get that
\begin{equation}\label{phi'+f}
\phi'+f=Cr+D/r^{2}.
\end{equation}
The $0ij$ component of Eq. (\ref{Lambda0-2ndEq2}) is an identity.
The $i0j$ component of Eq. (\ref{Lambda0-2ndEq2}) is
\begin{displaymath}
\partial^{d}\partial_{d}K_{j0i}-\partial^{d}\partial_{i}K_{j0d}=0,
\end{displaymath}
which gives
\begin{displaymath}
\delta_{ij}(-\triangle k-h'/r)+(x_{i}x_{j}/r^{2})[\triangle
k+h'/r-2k'/r-2(h+k)/r^{2}]=0,
\end{displaymath}
\begin{displaymath}
i.e.\quad\triangle k+h'/r=0, \quad \triangle
k+h'/r-2k'/r-2(h+k)/r^{2}=0,
\end{displaymath}
or equivalently
\begin{equation}\label{k'}
h+k+rk'=0.
\end{equation}
The $ijk$ component of Eq. (\ref{Lambda0-2ndEq2}) is
\begin{displaymath}
\partial^{d}\partial_{d}\partial_{k}\gamma_{ji}-
\partial^{d}\partial_{d}\partial_{j}\gamma_{ki}-
\partial^{d}\partial_{i}\partial_{k}\gamma_{jd}+
\partial^{d}\partial_{i}\partial_{j}\gamma_{kd}+
2\partial^{d}\partial_{d}K_{kji}-2\partial^{d}\partial_{i}K_{kjd}=0,
\end{displaymath}
which gives
\begin{displaymath}
\triangle[(2\psi/r^{2}-2g/r)(\delta_{ik}x_{j}-\delta_{ij}x_{k})]=0,
\end{displaymath}
\begin{displaymath}
i.e.\quad (2\psi/r^{2}-2g/r)''+(4/r)(2\psi/r^{2}-2g/r)'=0.
\end{displaymath}
Solving this equation we get that
\begin{equation}\label{Lambda0-psi}
\psi/r^{2}-g/r=B/r^{3}+A.
\end{equation}

From Eqs. (\ref{Curvature}), (\ref{1stordermetric2}) and
(\ref{staticO3torsion2}) the components of $R_{abcd}$ in
$\{x^{\mu}\}$ could be attained as follows:
\begin{equation}
\left\{\begin{array}{ll}R_{0i0j}=(\phi''+f')x_{i}x_{j}/r^{2}+(\phi'+f)(\delta_{ij}r^{2}-x_{i}x_{j})/r^{3},\\
R_{0ijk}=0,\\
R_{ijk0}=(2/r^{2})(h+k+rk')x_{[i}\delta_{j]k},\\
R_{ijkl}=(\psi/r^{2}-g/r)'(2/r)x_{[i}(\delta_{j]k}x_{l}-\delta_{j]l}x_{k})
-4(\psi/r^{2}-g/r)\delta_{i[k}\delta_{l]j}.
\end{array}\right.
\end{equation}
Substituting Eqs. (\ref{phi'+f}), (\ref{k'}) and (\ref{Lambda0-psi})
into the above equation, we get that
\begin{equation}\label{reducedRiemann}
\left\{\begin{array}{ll}R_{0i0j}=(Cr+D/r^{2})'x_{i}x_{j}/r^{2}+(Cr+D/r^{2})(\delta_{ij}r^{2}-x_{i}x_{j})/r^{3},\\
R_{0ijk}=0,\\
R_{ijk0}=0,\\
R_{ijkl}=(B/r^{3}+A)'(2/r)x_{[i}(\delta_{j]k}x_{l}-\delta_{j]l}x_{k})
-4(B/r^{3}+A)\delta_{i[k}\delta_{l]j}.
\end{array}\right.
\end{equation}
Then
\begin{equation}\label{sym-tracefree}
\begin{array}{ll}
|R|^{2}\eta_{00}-4R_{0cde}R_{0}{}^{cde}=12(C^{2}-4A^{2})+24(D^{2}-B^{2})/r^{6},\\
|R|^{2}\eta_{0i}-4R_{0cde}R_{i}{}^{cde}=0,\\
|R|^{2}\eta_{ij}-4R_{icde}R_{j}{}^{cde}=\\
\qquad \qquad \quad (4C^{2}-16A^{2}-16CD/r^{3}-32AB/r^{3}+16D^{2}/r^{6}-16B^{2}/r^{6})\delta_{ij}\\
\qquad \qquad \quad +(48CD/r^{3}+96AB/r^{3}
-24D^{2}/r^{6}+24B^{2}/r^{6})x_{i}x_{j}/r^{2},
\end{array}
\end{equation}
\begin{equation}\label{Ricci}
\begin{array}{l}
R_{00}=3C,\\
R_{0i}=R_{i0}=0,\\
R_{ij}=-(C+D/r^{3}+4A+B/r^{3})\delta_{ij}+[3(B+D)/r^{3}]x_{i}x_{j}/r^{2},\\
\ R\ =-6(C+2A),
\end{array}
\end{equation}
Applying Eq. (\ref{sym-tracefree}), the symmetric trace-free
equation (\ref{Lambda0SymEq}) could be solved, resulting in the
following relations:
\begin{equation}\label{CDAB}
C=\pm 2A,\ B=\pm D,\ CD+2AB=0.
\end{equation}

From Eq. (\ref{Ricci}) one could see that $R_{[ab]}=0$ and thus the
antisymmetric equation (\ref{Lambda0antisymEq}) gives
$\partial_{c}S_{[ab]}{}^{c}=0$. Components of
$\partial_{c}S_{ab}{}^{c}$ are as follows:
\begin{equation}\label{derivativetorsion}
\begin{array}{ll}\partial_{c}S_{00}{}^{c}=-f'-2f/r,\\
\partial_{c}S_{0i}{}^{c}=0,\\
\partial_{c}S_{i0}{}^{c}=x_{i}[h'/r+2(h+k)/r^{2}],\\
\partial_{c}S_{ij}{}^{c}=[-2(g/r)-r(g/r)']\delta_{ij}+(g/r)'x_{i}x_{j}/r.
\end{array}
\end{equation}
Therefore, $\partial_{c}S_{[ab]}{}^{c}=0$ results in
\begin{equation}\label{h'}
h'/r+2(h+k)/r^{2}=0.
\end{equation}
Combining Eqs. (\ref{k'}) and (\ref{h'}) we have
\begin{displaymath}
(h-2k)'=0,\quad h+k=-r(h+k)'/3,
\end{displaymath}
\begin{equation}\label{h-k}
h=2k+C_{1},\quad h+k=C_{2}/r^{3}.
\end{equation}

Now we turn to the trace equation (\ref{Lambda0TraceEq}). From Eqs.
(\ref{derivativetorsion}), (\ref{phi'+f}) and (\ref{Lambda0-psi}) we
have
\begin{eqnarray}\label{derivativetorsion2}
\partial_{c}S^{bc}{}_{b}&=&-f'-2f/r+2r(g/r)'+6(g/r)
\nonumber\\
&=&2\psi'/r+2\psi/r^{2}+\triangle\phi-3(2A+C).
\end{eqnarray}
Substituting Eqs. (\ref{Ricci}) and (\ref{derivativetorsion2}) into
Eq. (\ref{Lambda0TraceEq}) results in
\begin{equation}\label{Lambda0-phi-psi-Eq}
-(\triangle\phi+2\psi'/r+2\psi/r^{2})+9(2A+C)=(l^{2}/8\kappa)\rho.
\end{equation}
Equation. (\ref{Lambda0-phi-psi-Eq}) is an underdetermined equation
for $\phi$ and $\psi$. To solve it, more conditions are needed.
Suppose that the external solution is the weak Schwarzschild field,
i.e., $\phi=\psi=-GM/r$ and the internal solution is regular and
satisfies $\triangle\phi=4\pi G\rho$. Also, it is assumed that the
internal solution could be linked to the external solution smoothly
at $r=R_{S}$, resulting in a complete solution. For this complete
solution, there would be $C=-2A$,
\begin{equation}\label{Lambda0phi}
\phi=\int_{0}^{r}[Gm(r)/r^{2}]dr+\phi(0)
\end{equation}
with
\begin{displaymath}
m(r)=\int_{\mathbb{B}^{3}(r)}\rho=\int_{0}^{r}4\pi\rho r^{2}dr,
\end{displaymath}
\begin{displaymath}
\phi(0)=-GM/R_{S}-\int_{0}^{R_{S}}[Gm(r)/r^{2}]dr,
\end{displaymath}
and
\begin{displaymath}
\psi=[-Gm(r)/r](l^{2}/64\pi G\kappa+1/2).
\end{displaymath}
Note that $\psi(R_{S})=-Gm(R_{S})/R_{S}$, thus,
\begin{equation}\label{Lambda0psi}
\psi=-Gm(r)/r,
\end{equation}
\begin{equation}\label{Lambda0kappa}
\kappa=l^{2}/32\pi G.
\end{equation}
Equation. (\ref{Lambda0psi}) is just the same as the corresponding
case in GR. The result $C=-2A$ is in accordance with Eq.
(\ref{CDAB}). The torsion solutions will be given later in the more
general case.

Actually, Eq. (\ref{Lambda0-phi-psi-Eq}) can also be solved with
other supplementary conditions. For example, instead of assuming
$\triangle\phi=4\pi G\rho$, we may let
\begin{equation}\label{supplementary}
\triangle\phi=4\pi G(\rho+\tilde{\rho}),
\end{equation}
where $\tilde{\rho}$ is an arbitrarily given function and does not
contribute to the energy-momentum-stress tensors of matter fields.
Fixing $\kappa$ by Eq. (\ref{Lambda0kappa}), then from Eqs.
(\ref{CDAB}), (\ref{Lambda0-phi-psi-Eq}) and (\ref{supplementary})
we have
\begin{equation}\label{alternative-phi}
\phi=G\int_0^r \{[m(r)+\tilde{m}(r)]/r^{2}\}dr+\phi(0),
\end{equation}
\begin{equation}\label{alternative-psi}
\psi=-\frac{G}{r}[m(r)+\tilde{m}(r)/2]+\frac{3}{2}(2A+C)r^2
\end{equation}
with
\begin{displaymath}
\tilde{m}(r)=\int_{\mathbb{B}^{3}(r)}\tilde{\rho}
=\int_{0}^{r}4\pi\tilde{\rho}r^{2}dr,\quad C=\pm 2A.
\end{displaymath}
As the internal solutions are regular, from Eqs. (\ref{phi'+f}),
(\ref{Lambda0-psi}) and (\ref{h-k}), there should be $B=D=0$,
$C_2=0$. Therefore, the torsion solutions corresponding to Eqs.
(\ref{alternative-phi}) and (\ref{alternative-psi}) are as follows:
\begin{equation}\label{f}
f=Cr-G[m(r)+\tilde{m}(r)]/r^{2},
\end{equation}
\begin{equation}\label{g}
g=(4A+3C)r/2-G[m(r)+\tilde{m}(r)/2]/r^{2},
\end{equation}
\begin{equation}
h=-k=\frac{1}{3}C_1.
\end{equation}

The former case which is compatible with the Schwarzschild solutions
corresponds to the special choice with $C=-2A$ and $\tilde{\rho}=0$.
In fact, another choice of $\tilde{\rho}$ could deduce the galactic
rotation curves without invoking dark matter. To fit the galactic
rotation curves, the dark matter density profile $\rho_{DM}$ has
been given for many spiral galaxies, for example, see Refs.
\cite{DM1,DM2}, where the following choice is made:
\begin{equation}\label{halo}
\rho_{DM}=\frac{\sigma^{2}}{2\pi G(r^{2}+a^{2})}.
\end{equation}
In Refs. \cite{DM1,DM2}, the mass distribution of the galaxies is
modeled as the sum of the bulge and disk stellar components and a
halo of dark matter. The rotation curves are used to determine the
two halo parameters $\sigma$ and $a$. In our model, we may just let
the gravitational contribution $\tilde \rho$ to take the same form
as the dark matter density profile and suitably choose the
integration constants, i.e., $\tilde{\rho}=\rho_{DM}$ and $C=-2A$,
without the inclusion of any real dark matter. For this case, Eqs.
(\ref{alternative-phi}) and (\ref{alternative-psi}) is almost
equivalent to the corresponding case in GR with dark matter. The
parameters $\sigma$ and $a$ can be determined in the same way as
that in Refs. \cite{DM1,DM2} and, therefore, with the same values.
For example, for the galaxy NGC 2841, $\sigma=232$ km/s and $a=11.6$
kpc fit the rotation curve well, for the galaxy NGC 3031,
$\sigma=86$ km/s and $a=2.0$ kpc fit the rotation curve well, and so
on. To say `almost equivalent' but not `equivalent', is because the
term $\tilde{m}/2$ in Eq. (\ref{alternative-psi}) is different from
$\tilde{m}$, which should be the case in GR with dark matter.
Fortunately, $\psi$ would not affect the rotation curves in the
leading order of approximation, since the rotation velocity $v_{c}$
in a galaxy at a radius $r$ in the approximation is given by
\begin{equation}
v_{c}^2 (r)=r(\partial \phi/\partial r).
\end{equation}
For higher order approximations, more work is needed to be done and
experiments with higher accuracy are needed to check the
corresponding results.

One may argue that $\tilde{\rho}$ can not be specified $a\ priori$
and could only be determined from observation. Actually, such kinds
of functions also appear in other models which attempt to explain
the galactic rotation curves without involving dark matter, such as
Milgrom's modification of Newtonian dynamics (MOND)
\cite{MOND1,MOND2} and the modified Newton's gravity in Finsler
space \cite{finsler}. What we can do now is to point out the
geometrical meaning of $\tilde{\rho}$. From Eqs. (\ref{f}) and
(\ref{g}), there is the relation:
\begin{equation}\label{rhotilde}
4\pi G\tilde{\rho}=-\frac{2}{r^{2}}[r^{2}(f-g)]'-3(4A+C),
\end{equation}
which shows that the dark matter density could be directly related
to the spacetime torsion.

When $l\rightarrow\infty$, $l^2/\kappa$ should tend to a finite
value, as was mentioned before. Obviously Eq. (\ref{Lambda0kappa})
satisfies this requirement. In \cite{Wu,An,Guo79,Guo07}, the
coupling constant is chosen to be $-l^{2}/64\pi G$, which is
different from Eq. (\ref{Lambda0kappa}). For this case, there exists
no regular internal solution which satisfies $\triangle\phi=4\pi
G\rho$ and has a smooth junction to the Schwarzschild solutions. In
fact, with this choice, the ratio of the coefficient of the Einstein
term $G_{ab}$ to that of the matter term $\Sigma_{ab}$ in Eq.
(\ref{1stEq}) would be $1:(-8\pi G)$, just like the case in GR. But
the role of the Einstein term in our model is different from that of
GR. Actually, from Eq. (\ref{Ricci}), the Einstein term only
contributes a constant term to Eq. (\ref{Lambda0TraceEq}), while the
torsion term plays an important role in that equation.

Generally, the torsion tensor can be decomposed \cite{Hayashi1,Hehl}
into three irreducible parts with respect to the Lorentz group: the
tensor part, trace-vector part, and the axial vector part. For
static and $O(3)$-symmetric torsion, the axial vector part vanishes
automatically, the tensor part satisfies $f=2g$, $h=2k$ and the
trace-vector part satisfies $f=-g$, $h=-k$. By Eqs. (\ref{f}) and
(\ref{g}), if the torsion field only contains the tensor part, the
matter density should be a constant:
\begin{equation}\label{tensor-rho}
\rho=3(2A+C)/2\pi G.
\end{equation}
If the torsion field only contains the trace-vector part, then
\begin{equation}\label{trace-vector-rho}
4\rho+3\tilde{\rho}=3(4A+5C)/4\pi G.
\end{equation}
When $\tilde{\rho}=0$, the matter density has to be a constant, too.

\section{Remarks}
The weak field approximation of the $\Lambda\rightarrow 0$ limit of
the dS gravity model has been calculated in the static and
spherically symmetric case. The matter field is assumed to be a
smoothly distributed dust sphere with finite radius. It comes out
that if and only if $\kappa=l^{2}/32\pi G$, there exist regular
internal solutions which satisfy $\triangle\phi=4\pi G\rho$ and have
a smooth junction to the weak Schwarzschild fields. Recall that the
main part of the action is the quadratic curvature term, not the
Einstein--Hilbert term. The existence of the above solutions is of
significance. The choice of the coupling constant here is different
from that of \cite{Wu,An,Guo79,Guo07}, where $\kappa=-l^{2}/64\pi
G$. The choice in \cite{Wu,An,Guo79,Guo07} may due to a comparison
between the dS gravity model and GR. But the Einstein term in the dS
gravity model plays a different role from that of GR, as it only
contributes to a constant term in the weak field approximate
equations. Actually, the metric components and torsion components
are closely related by Eqs. (\ref{phi'+f}) and (\ref{Lambda0-psi}),
such that the curvature tensor has to take a special form
(\ref{reducedRiemann}). Moreover, one may let $C=A=0$ and $B=D=0$,
then $R_{abcd}=0$, i.e., the Weizenb\"ock spacetime
\cite{Weitzenbock1,Weitzenbock2} would be attained. On the other
hand, the torsion tensor plays an important role in this model. In
the weak field approximate solutions, if the torsion tensor only
contains the tensor part, the matter density should be a constant;
if it only contains the trace-vector part, Eq.
(\ref{trace-vector-rho}) should be upheld. In particular, the matter
density should be a constant in the torsion-free case.

The trace equation (\ref{Lambda0TraceEq}) results in an
underdetermined equation for the metric components $\phi$ and $\psi$
in the weak field approximation. Solutions with $\triangle\phi=4\pi
G(\rho+\tilde{\rho})$ could be attained, which can explain the
galactic rotation curves without the help of introducing dark
matter. The geometrical meaning of $\tilde{\rho}$ has been given by
Eq. (\ref{rhotilde}). It is a geometric quantity related to the
spacetime torsion and can play the role of dark matter density
though it is irrelevant to the energy-momentum-stress tensors of
matter fields. To see the higher order behavior of the model, more
work is needed to be done and experiments with higher accuracy are
needed to check the corresponding results.

Finally, it should be remarked that all the above results need to be
reexamined in the case with $\Lambda \neq 0$. But we can conclude,
at least, that if there exist regular internal solutions which are
in accordance with the Newtonian gravitational law and could be
smoothly extended to the weak S-dS fields, the coupling constant
should be chosen as $\kappa=l^2/32\pi G$.

\phantomsection \addcontentsline{toc}{section}{Acknowledgments}
\section*{Acknowledgments}
One of us (Lu) would like to thank Prof. Zhi-Bing Li, Xi-Ping Zhu,
and thank the late Prof. Han-Ying Guo for their sincere help. He
also expresses his appreciation for hospitality during his stay at
the Institute of High Energy Physics, and Theoretical Physics Center
for Science Facilities, Chinese Academy of Sciences. This work is
supported by the National Natural Science Foundation of China under
Grant Nos. 10975141, 10831008, 11275207, and the oriented projects
of CAS under Grant No. KJCX2-EW-W01.

\phantomsection 
\end{document}